\documentclass[12pt]{iopart}
\usepackage{graphicx}
\begin{document}
\date{}

\title[New aspects of integrability]{New aspects of integrability of force-free 
Duffing-van der Pol oscillator and related nonlinear systems}

\author{V~K~Chandrasekar, M~Senthilvelan and M~Lakshmanan}

\address{Centre for Nonlinear Dynamics, Department of Physics,  Bharathidasan
University, Tiruchirapalli - 620 024, India}

\begin{abstract} 

In this letter we show that the force-free Duffing-van der Pol oscillator 
is completely integrable for a specific parametric choice. We derive a 
general solution for this parametric choice. Further, we describe a 
procedure to construct the transformation which removes the time dependent 
part from the first integral and provide the general solution by quadrature. 
The procedure is shown to have a wider applicability through 
additional examples. We also show 
that through our method one can deduce linearizing transformations in a 
simple and straightforward way and illustrate it with a specific example.
\end{abstract}

%\pacs{}
\section{Introduction}
One of the well-studied but still challenging equations in nonlinear dynamics
is the Duffing-van der Pol oscillator equation \cite{ref1}. Its autonomous version 
(force-free) is
\begin{eqnarray} 
\ddot{x}+(\alpha+\beta x^2)\dot{x}-\gamma x +x^3=0, \label {eq1}
\end{eqnarray}
where  over dot denotes differentiation with respect to time and 
$\alpha, \beta$ and $ \gamma$ are arbitrary parameters. Equation (\ref {eq1})
arises in a model describing the propagation of voltage pulses along a 
neuronal axon and has received a lot of attention recently by many authors.  
A vast amount of literature exists on this equation, for
details see for example \cite {ref1,ref2} and references therein.
In most of the works non-integrability properties have been considered
primarily since the forced version of equation (\ref {eq1}) exhibits a rich variety 
of bifurcations and chaotic phenomena. When $\beta=0$, equation (\ref {eq1}) 
becomes the (force-free) Duffing oscillator whose integrability and non-integrability 
properties are well-known \cite {ref1}, while when the cubic term is absent 
it is the standard van der Pol equation.

As far as the integrability properties of equation (\ref {eq1}) are 
concerned not much progress has been made mainly due to the fact that 
it does not pass the Painlev\'{e} test as it admits a movable 
algebraic branch point and a local Laurent expansion in the form 
\cite{ref3}
\begin{eqnarray} 
x(t)=\sqrt{\frac {3}{2\beta}}\tau^{-\frac {1}{2}}+
\sqrt{\frac {3}{2\beta}}\left(\frac {3}{2\beta}-\frac{\alpha}{2}\right)
\tau^{\frac {1}{2}}+
a_3\tau+\ldots, \label {eq2}
\end{eqnarray}
where $\tau=(t-t_0)$ and $t_0$ and $a_3$ are arbitrary constants.
However, it is known that the system (\ref{eq1}) admits nontrivial symmetries 
and a first integral for a specific parametric choice \cite{ref4}
\begin{eqnarray} 
\alpha=\frac {4}{\beta},\;\;\;\;\; \gamma=-\frac{3}{\beta^2}.\label{eq3}
\end{eqnarray}
The associated first integral reads  
\begin{eqnarray} 
e^{\frac {3}{\beta}t}\left(\dot {x}+\frac {1}{\beta}x
+\frac {\beta}{3}x^3\right)=I. \label {eq4}
\end{eqnarray}
Rewriting (\ref {eq4}) we get
\begin{eqnarray} 
\dot {x}+\frac {1}{\beta}x
+\frac {\beta}{3}x^3=Ie^{-\frac {3}{\beta}t}, \label {eq4a}
\end{eqnarray}
which is nothing but a special case of the Abel's equation of the first kind 
\cite {ref5}.However, 
it has not been explicitly integrated directly due to the 
explicit time dependent part in the first integral and so the complete 
integrability of (\ref {eq1}) for the choice (\ref {eq3}) remains unclear 
till date. 
In this work, we integrate the equation (\ref {eq1})
explicitly for the parametric choice (\ref {eq3}) and provide a 
general solution for the above specific parametric choice of the 
Duffing-van der Pol nonlinear oscillator, 
thereby establishing its complete integrability. Further, we show that our
procedure has a wider applicability, including linearization, through additional
examples.
\section {Integration of Duffing-van der Pol oscillator for the 
specific parametric choice (\ref {eq3})}
Rewriting (\ref {eq1}) with the specific parametric choice (\ref {eq3}) we 
get
\begin{eqnarray} 
\ddot{x}+\left(\frac {4}{\beta}+\beta x^2\right)\dot{x}+\frac{3}{\beta^2}x 
+x^3=0. \label{eq5}
\end{eqnarray}
Now introducing a transformation
\begin{eqnarray} 
w= -x e^{\frac{1}{\beta}t},\;\;\;\;\; 
z= e^{-\frac{2}{\beta}t},\label {eq6}
\end{eqnarray}
where $w$ and $z$ are new dependent and independent variables respectively,  
equation (\ref {eq5}) can be transformed to
\begin{eqnarray} 
w''-\frac {\beta^2}{2}w^2w'=0,
\label {eq7a}
\end{eqnarray}
where prime denotes differentiation with respect to $z$. Equation (\ref{eq7a}) 
can be integrated trivially to yield 
\begin{eqnarray}
w'-\frac{\beta^2}{6}w^3=I, 
\label {eq7}
\end{eqnarray}
where $I$ is the integration constant. Equivalently, the transformation 
(\ref {eq6}) reduces (\ref {eq4}) to this form. Solving (\ref {eq7}),
we obtain 
\begin{eqnarray} 
z-z_0=\frac {a}{3I}\left[\frac{1}{2}
\log\left(\frac{(w+a)^2}{w^2-aw+a^2}\right)
+\sqrt{3}\,\mbox{arctan}\left(\frac{w\sqrt{3}}{2a-w}\right)\right], \label {eq8}
\end{eqnarray}
where $a=\sqrt[3]{\frac {6I}{\beta ^2}}$
and $z_0$ is the second integration constant (see figure 1 for a plot of the
solution $x(t)$). 
Replacing the second term inside the square bracket in (\ref {eq8})
in terms of logarithmic function we get 
\begin{eqnarray} 
\fl \quad \qquad 
\frac {1}{2}\left[i\sqrt{3}\log\left(\frac{-2+\bar{w}+i\sqrt{3}\bar{w}}
{-2+\bar{w}-i\sqrt{3}\bar{w}}\right)+\log\left(\frac{a(\bar{w}+1)^2}
{(\bar{w}^2-\bar{w}+1)}\right)\right]=\frac {3I}{a}(z-z_0),
\label {eq9}
\end{eqnarray} 
where $ \bar{w}=\frac {w}{a}$. This can be further simplified to 
\begin{eqnarray} 
\left[\frac{-2+\bar{w}(1+i\sqrt{3})}{-2+\bar{w}(1-i\sqrt{3})}\right]^{i\sqrt{3}}
\left[\frac{\bar{w}+1}{(\bar{w}^2-\bar{w}+1)}\right]=\frac{1}{a}
e^{\frac {6I}{a}(z-z_0)}.\label {eq10}
\end{eqnarray}
Rewriting $\bar{w}$ and $z$ in terms of old variables one can get the 
explicit solution for the equation (\ref {eq5}).
\begin{figure}[!ht]
\centering{\includegraphics[width=0.6\linewidth]{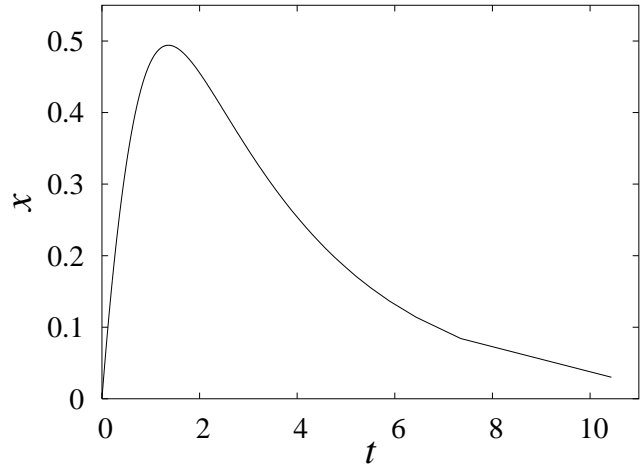}}
\caption{The solution plot of the Duffing-van der Pol oscillator 
equation (\ref{eq5}) in the form (\ref{eq8}) when $\beta=3$.}
\label{fig:plot6}
\end{figure}

  One may note that a simple solution for equation (\ref {eq5}) can also
be built from the first integral (\ref {eq7}). Restricting $I=0$ 
in equation (\ref {eq7}) yields
\begin{eqnarray} 
w=\frac{\sqrt {3}}{\beta }(z_0-z)^{-\frac {1}{2}}.\label {eq10a}
\end{eqnarray}
Rewriting (\ref {eq10a}) in terms of the old coordinates we get a particular 
solution for (\ref {eq5}) in the form
\begin{eqnarray} 
x=-\frac {\sqrt{3}}{\beta\sqrt{(t_0e^{\frac{2}{\beta}t}-1)}}, \qquad t_0>1.
\end{eqnarray}
\section{Method of constructing transformation} 
 One can construct the transformation (\ref {eq6}) systematically from 
the first integral (\ref {eq4}). For example, let us split the 
functional form of the first integral $I$ into two terms such that 
one involves all the variables 
$(t,x,\dot{x})$ while the other excludes $\dot{x}$, that is, 
\begin{eqnarray} 
I=F_1(t,x,\dot{x})+F_2(t,x). \label {eq11}
\end{eqnarray}
Now let us split 
the function $F_1$ further in terms of two functions such that $F_1$ itself 
is a function of the product of the two functions, say, a perfect differentiable function
 $\frac {d}{dt}G_1(t,x)$ and another function $G_2(t,x,\dot{x})$, that is,
\begin{eqnarray} 
I=F_1\left(\frac{1}{G_2(t,x,\dot{x})}\frac {d}{dt}G_1(t,x)\right)
+F_2\left(G_1(t,x)\right).
\label {eq12}
\end{eqnarray}
We note that while rewriting equation (\ref {eq11}) in the form (\ref {eq12}), we
demand the function $F_2(t,x)$ in (\ref {eq11}) automatically to be a 
 function of $G_1(t,x)$. Now identifying  the function  $G_1$ as the 
new dependent variable and the integral of $G_2$ over time as the new 
independent variable, that is,
\begin{eqnarray} 
w=G_1(t,x),\nonumber\\ 
z=\int_o^t G_2(t',x,\dot{x}) dt', \label {eq13}
\end{eqnarray}
one indeed obtains an explicit transformation to remove  the time dependent 
part in the first integral (\ref {eq4}). We note here that the integration 
on the
right hand side of (\ref {eq13}) leading to $z$ can be performed provided the 
function $G_2$ is an exact derivative of $t$, that is,
$G_2=\frac{d}{dt}z(t,x)=\dot{x}z_x+z_t$, so that $z$ turns out to be a function
$t$ and $x$ alone. In terms of the new variables,
equation (\ref {eq12}) can be modified to the form
\begin{eqnarray} 
I=F_1\left(\frac {dw}{dz}\right)+F_2(w).\nonumber
\end{eqnarray}
In other words
\begin{eqnarray} 
F_1\left(\frac {dw}{dz}\right)=I-F_2(w).\label {eq13a}
\end{eqnarray}
Now rewriting equation (\ref {eq13a}) one obtains a separable equation 
\begin{eqnarray}  
\frac {dw}{dz}=f(w),
\end{eqnarray}
which can be integrated by quadrature.

For the special case of the Duffing-van der Pol oscillator equation (\ref
{eq5}), by following the procedure given above, we can deduce the
transformation (\ref {eq6}) systematically from the first integral 
(\ref {eq4}). For this purpose, we rewrite equation (\ref {eq4a}) as
\begin{eqnarray} 
\hat{I}=\frac{\beta}{2}\left(\dot {x}+\frac {1}{\beta}x\right)e^{\frac {3}{\beta}t}
+\frac {\beta^2}{6}x^3e^{\frac {3}{\beta}t}, \label {eq14}
\end{eqnarray}  
where $\hat{I}=\frac{\beta}{2}I$. Equation (\ref {eq14}) can be further split into
\begin{eqnarray} 
\hat{I}=-\frac{\beta}{2}e^{\frac {2}{\beta}t}\frac {d}{dt}\left(-xe^{\frac {1}{\beta}t}\right)
+\frac {\beta^2}{6}\left(xe^{\frac {1}{\beta}t}\right)^3. \label {eq14a}
\end{eqnarray}
Comparing equation (\ref {eq14a}) with (\ref {eq12}) we get
\begin{eqnarray} 
G_1= -x e^{\frac{1}{\beta}t},\;\;\;\;\; 
G_2= \frac{2}{\beta}e^{-\frac{2}{\beta}t},
\end{eqnarray}
which in turn provides the transformation coordinates, $w$ and $z$, through the
relation (\ref {eq13}) of the form
\begin{eqnarray} 
w= -x e^{\frac{1}{\beta}t},\;\;\;\;\; 
z= e^{-\frac{2}{\beta}t}.\nonumber
\end{eqnarray}
\section{Applications} 
 We observe that the above procedure can be extended to solve a class of 
equations. In the following we briefly describe the applicability of this 
method to Duffing oscillator and the equation describing the motion of a 
gaseous general-relativistic fluid sphere. A detailed connection between 
the integrating factors, integrals of motion and the transformation 
coordinates will be published elsewhere \cite{ref6}.
\subsection{Duffing oscillator}
 Let us consider the Duffing oscillator \cite {ref1}
\begin{eqnarray}	   
\ddot{x}+\alpha \dot{x}+\beta x+x^3=\gamma cos\omega t,   \label {eq15}
\end{eqnarray}
where $\alpha, \beta, \gamma$ and $\omega$ are arbitrary parameters. 
In the absence of the external forcing $(\gamma=0)$, equation (\ref {eq15}) 
has been shown to be integrable for a specific parametric choice \cite {ref7}, 
namely, $2\alpha^2=9\beta$. The first integral for this parametric 
choice has been constructed in the form
\begin{eqnarray}
I=e^{\frac{4}{3}\alpha t} \bigg[\frac{\dot{x}^2}{2}+\frac{\alpha x\dot{x}}{3}
+\frac{\alpha^2 x^2}{18}+\frac{x^4}{4}\bigg].\label {eq16}
\end{eqnarray}
Equation (\ref{eq16}) is a complicated nonlinear first order ordinary differential 
equation (ODE) with explicit time dependent coefficients and so cannot be integrated 
directly. A usual way to overcome this problem is to transform 
equation (\ref{eq16}) to an autonomous equation and integrate it. Using 
our procedure one can construct the required transformations systematically. 
For example, rewriting equation (\ref{eq16}) in the form (\ref{eq11}) we get
\begin{eqnarray}
I=\frac{1}{2}\left(\dot{x}+\frac{\alpha x}{3}\right)^2 e^{\frac{4}{3}\alpha t}
+\frac{x^4}{4}e^{\frac{4}{3}\alpha t}. \label {eq17}
\end{eqnarray}
Now splitting the first term in equation (\ref{eq17}) further in the form of 
(\ref{eq12}),
\begin{eqnarray} 
I=\bigg[e^{\frac {\alpha}{3}t}\frac{d}{dt}\left(\frac{x}{\sqrt{2}}
e^{\frac{\alpha}{3}t}\right)\bigg]^2
+\left(\frac{x}{\sqrt{2}} e^{\frac {\alpha}{3}t}\right)^4, \label {eq18}
\end{eqnarray}
and identifying the dependent and independent variables from (\ref{eq18})
using the relations (\ref {eq13}), we obtain the transformation
\begin{eqnarray} 
w=\frac{1}{\sqrt{2}}x e^{\frac {\alpha t}{3}},\;\;\;\;
z=-\frac {3}{\alpha}e^{-\frac {\alpha t}{3}}. \label {eq19}
\end{eqnarray}
One can easily check that equation (\ref{eq16}) can be transformed into the autonomous 
form with the help of the transformation (\ref{eq19}) and the latter can be
integrated in terms of Jacobian elliptic function \cite {ref7}. We note that the
transformation (\ref{eq19}) exactly coincides with the one known in the 
literature which has been constructed in an ad hoc way.
\subsection{Static gaseous general relativistic fluid sphere }
Recently, Duarte etal \cite {ref8} have constructed the first integral for an 
equation describing the static gaseous general relativistic fluid sphere 
\cite {ref9}, 
\begin{eqnarray}
\ddot{x}=\frac {t^2\dot{x}^2+x^2-1}{t^2x}, \label {eq20}
\end{eqnarray}
through the so called Prelle-Singer procedure, in the form 
\begin{eqnarray}
I=\frac {2tx\dot{x}+x^2+t^2\dot{x}^2-1}{t^2x^2}. \label {eq21}
\end{eqnarray}
We note that unlike the earlier two examples, namely Duffing and Duffing- 
van der Pol oscillators, we have no exponential function in the first integral.
Also the first integral is a rational function. However, in the
following we show that one can integrate it and obtain a general solution for
this problem. By following our procedure, equation (\ref{eq21}) 
can be recast in the form 
\begin{eqnarray}
I=\left(\frac {x+t\dot{x}}{tx}\right)^2-\frac{1}{t^2x^2}. \label {eq21a}
\end{eqnarray}
One may note that the first term in (\ref{eq21a}) can be written in terms of a 
perfect differential form, that is, $(\frac{d}{dt}[\log(tx)])^2$ so that 
$G_1=\log(tx)$ and the function $G_2$ turns out to be a constant,  $G_2=1$,
 in the present example. As a consequence we obtain a transformation
\begin{eqnarray}
w=\log(tx),\qquad z=t, \label {eq22}
\end{eqnarray}
which can be utilized to rewrite equation (\ref{eq21}) in the autonomous form
\begin{eqnarray}
w'^2-\exp(-2w)=I.\label {eq23}
\end{eqnarray} 
Equation (\ref{eq23}) can be integrated under two different choices \cite{ref10}, 
namely, $(i)  I>0$ and $(ii) I<0$. The respective solutions are
\begin{eqnarray} 
(i)\;\;\;\;\; z-z_0=-\frac {1}{2\sqrt{I}}\log\left(\frac{\sqrt{I+e^{-2w}}
-\sqrt{I}}{\sqrt{I+e^{-2w}}+\sqrt{I}}\right),\;\;\;\;\;   I>0,\nonumber\\
(ii)\;\;\;\;\; z-z_0=-\frac {1}{\sqrt{-I}}\arctan\left(\frac{\sqrt{I+e^{-2w}}}
{\sqrt{-I}}\right),\;\;\;\;\;\;   I<0,\label {eq23b}
\end{eqnarray}
where $z_0$ is the second integration constant. Rewriting equation (\ref{eq23b}) 
we get
\begin{equation}
w=\left\{
\begin{array}{ll}
 {\log\left(\displaystyle{\frac{1-e^{-2\sqrt{I}(z-z_0)}}
 {2\sqrt{I}e^{-\sqrt{I}(z-z_0)}}}\right)}, &  I>0, \\
 {\log\left(\frac{\cos\sqrt{-I}(z-z_0)}{\sqrt{-I}}\right)},& I<0.
	 \label{eq23d}
\end{array}
\right.
\end{equation}
Utilizing (\ref{eq22}) in (\ref{eq23d}) one can write down the solution 
for the static gaseous general relativistic fluid sphere of the form
\begin{eqnarray} 
x=\left\{
\begin{array}{ll}
\displaystyle{\frac
{1}{\sqrt{I}t}}\displaystyle{\sinh\sqrt{I}(t-t_0)},&   I>0, \\
\displaystyle{\frac{1}{\sqrt{-I}t}}\displaystyle{
\cos\sqrt{-I}(t-t_0)},
& I<0.
\end{array}\right. \label {eq23f}
\end{eqnarray}
To our knowledge this solution is new to this problem.
\section{Linearization}
Interestingly, our method not only helps to integrate the first integral and
gives the complete solution for a class of second order nonlinear ODEs but it 
also helps to deduce the transformations which can be used to linearize the 
given nonlinear ODEs in a very systematic way. For example, let us 
consider the modified Emden equation,
\begin{eqnarray}	   
\ddot{x}+\alpha x\dot{x}+\beta x^3=0,   \label {eq24}
\end{eqnarray}
where $\alpha$ and $\beta$ are arbitrary parameters. Mahomed and 
Leach \cite {ref11} have shown that equation (\ref{eq24})
is linearizable through point transformations for a particular parametric 
choice, namely, $\alpha^2= 9\beta$.
The linearizing transformation and the first integral are known to be
\begin{eqnarray}
w=\frac{t}{x} -\frac{\alpha t^2}{6} ,\quad 
z=\frac{\alpha }{3}t-\frac{1}{x}, \label {eq25}
\end{eqnarray}
and
\begin{eqnarray}
I=-t+\frac{x}{\frac{\alpha }{3}x^2+\dot{x}} . \label {eq26} 
\end{eqnarray}
Substituting (\ref{eq25}) into (\ref{eq24}) one can transform the latter into the
free particle equation, that is, 
\begin{eqnarray}
\frac{d^2w}{dz^2}=0 . \label {eq27} 
\end{eqnarray}
The authors have derived the linearizing transformation
through Lie symmetry analysis \cite{ref11}.
However, in the following we show that the linearizing
transformation (\ref{eq25}) can also be derived from the first integral itself
through our method.

Rewriting the first integral (\ref{eq26}) in the form
\begin{eqnarray}
I=\frac{x^2}{\frac{\alpha }{3}x^2+\dot{x}}\left[\frac {d}{dt}\left(\frac{t}{x} 
-\frac{\alpha t^2}{6}\right)\right],
\label {eq28} 
\end{eqnarray}
and identifying (\ref{eq28}) with (\ref{eq12}) we get
\begin{eqnarray}
G_1=\frac{t}{x} -\frac{\alpha t^2}{6},\;\;\;\;\; 
G_2=\frac{\alpha }{3}+\frac{\dot{x}}{x^2},\;\;\;\;\;  F_2=0.
\end{eqnarray}
One can see that, unlike the earlier examples discussed so far, in the present
example $G_2$ depends both on $x$ and $\dot{x}$. However, as we noted earlier
the new variable $z$ can be obtained once the function $G_2$ is a perfect
derivative of $t$, that is,
\begin{eqnarray}
G_2=\frac{\alpha }{3}+\frac{\dot{x}}{x^2}\equiv \frac {d}{dt}
\left(\frac{\alpha }{3}t-\frac{1}{x}\right),
\end{eqnarray}
so that (\ref{eq13}) gives
\begin{eqnarray}
w=\frac{t}{x} -\frac{\alpha t^2}{6},\;\;\;\;\;  
z=\frac{\alpha }{3}t-\frac{1}{x}, \nonumber
\end{eqnarray}
which is nothing but the linearizing transformation. One may note that in this
case while rewriting the first integral $I$ (equation (\ref{eq26})) in the form 
(\ref{eq11}), the function  $F_2$ disappears, that is $F_2=0$, and as a 
consequence we arrive at (vide equation (\ref{eq13a})) 
\begin{eqnarray} 
\frac {dw}{dz}=I
\end{eqnarray}
which in turn gives (\ref{eq27}) by differentiation or leads to the 
solution by an integration. On the other hand vanishing of the function $F_2$ 
in this analysis is precisely the condition for the system to be transformed 
into the free particle equation \cite {ref6}.
\section{Conclusion}
 One of the common problems in solving second order nonlinear 
ODEs is how to solve the time dependent first integral and obtain a 
general solution associated with the given equation.
In fact most of the methods available to tackle the second order nonlinear 
ODEs provide first integrals only. In this work, we
have proposed a novel method to identify transformation coordinates which can
be used to rewrite the first integral without explicit time dependence so that
it can be integrated by quadrature. Interestingly, we showed that the
transformation coordinates can be constructed from the first integral itself.
With this choice, the above procedure can be augmented with other existing methods 
to obtain a complete solution for a given problem.
We have also shown the applicability of our method in constructing 
linearizing transformations as well in a simple and straightforward way. 
Finally, we mention that the proposed method 
can also be used to study certain coupled nonlinear oscillators, the details 
of which will be presented separately \cite {ref6}.
\\
\\
 The work of VKC is supported by CSIR in the form of a Junior Research
 Fellowship.  The work of MS and ML forms part of a Department of Science 
 and Technology, Government of India, sponsored research project.

\end{document}